\documentclass[11pt]{article}
\usepackage{epsfig,latexsym}

\newcommand{\sect}[1]{\setcounter{equation}{0}}

\input epsf

\begin{document}

\font\cmss=cmss10 \font\cmsss=cmss10 at 7pt 
\hfill CERN-TH/2000-139

\bigskip

\begin{center}
{\Large \textbf{LARGE-\textit{N} EXPANSION, CONFORMAL FIELD THEORY\\[0pt]
AND RENORMALIZATION-GROUP FLOWS\\[0pt]
\vskip 5pt IN\ THREE DIMENSIONS}}
\end{center}

\vspace{6pt}

\begin{center}
\textsl{Damiano Anselmi}

\textit{CERN, Theory Division, CH-1211, Geneva 23, Switzerland }
\end{center}

\vspace{20pt}

\begin{center}
\textbf{Abstract}
\end{center}

\vspace{4pt}{\small I study a class of interacting conformal field theories
and conformal windows in three dimensions, formulated using the Parisi large-%
}$N${\small \ approach and a modified dimensional-regularization technique.
Bosons are associated with composite operators and their propagators are
dynamically generated by fermion bubbles. Renormalization-group flows
between pairs of interacting fixed points satisfy a set of non-perturbative }%
$g\leftrightarrow 1/g${\small \ dualities. There is an exact relation
between the beta function and the anomalous dimension of the composite
boson. Non-Abelian gauge fields have a non-renormalized and quantized gauge
coupling, although no Chern--Simons term is present. A problem of the naive
dimensional-regularization technique for these theories is uncovered and
removed with a non-local, evanescent, non-renormalized kinetic term. The
models are expected to be a fruitful arena for the study of odd-dimensional
conformal field theory.}

\vskip 7.5truecm \noindent CERN-TH/2000-139 -- May, 2000

\vfill\eject 
\bigskip

Unexpected remnants of the renormalization algorithm in quantum field theory
are the Adler--Bell--Jackiw anomalies \cite{adler,jackiw}, finite amplitudes
arising from the quantum violation of classical conservation laws. Anomalies
fall into two main classes: axial and trace. The axial anomalies obey
all-order properties, such as the Adler--Bardeen theorem \cite{adlerbardeen}%
, and give important information about the low-energy physics, by means of
the 't Hooft anomaly-matching conditions \cite{thooftmatching}. The trace
anomaly is related to the beta function \cite{adler2}, by a formula $\Theta
=\beta ^{a}\mathcal{O}^{a}$, $\mathcal{O}^{a}$ being composite operators.
Certain trace anomalies in external fields can be computed exactly in the IR
limit of (supersymmetric) UV-free theories \cite{noi,noi2}, where an exact
beta function can also be derived \cite{nsvz}. They reveal the intrinsic
irreversibility of the renormalization-group flow, its relation to the
invariant area of the graph of the beta function between the fixed points 
\cite{athm} and the essential difference between marginal and relevant
deformations \cite{cea}.

Most of these powerful results apply only to even dimensions. Trace
anomalies in external gravitational and flavour fields do not exist in odd
dimensions. Nevertheless, an odd-dimensional formula for the irreversibility
of the RG\ flow can in principle be written \cite{at6d}, because the
relation $\Theta =\beta ^{a}\mathcal{O}^{a}$ is completely general and so is
the notion of invariant area of the graph of the beta function. It would be
desirable to dispose of a web of non-trivial conformal field theories,
conformal windows and RG\ flows in three dimensions to investigate these and
related issues more closely. The purpose of this letter is to construct a
large class of such theories and flows, and address the search for
appropriate odd-dimensional generalizations of the properties mentioned
above.

The beta functions of the most general power-counting renormalizable
three-dimensional theory with a Chern--Simons vector field have been studied
by Avdeev \textit{et al.} in ref. \cite{avdeev}. Three-dimensional quantum
field theory is relevant for its possible applications in the domain of
condensed-matter physics. However, the Chern--Simons models are
parity-violating and this somewhat limits the range of their applicability.
The three-dimensional 
$\phi^6$ theory is known to have a non-trivial fixed point in
the large-$N$ expansion and a conformal window interpolating
between the free limit and this point \cite{townsend}. 
Nevertheless, a large class of parity-preserving 
conformal windows is not known at present
and will be constructed here.

The Chern--Simons coupling $g_{\mathrm{cs}}$ is not renormalized \cite
{nonre,nonre2,kasputin}. The simplest argument to prove this fact proceeds as
follows. Let us denote by $\beta _{\mathrm{cs}}$ the beta function
of $g_{\mathrm{cs}}.$ The results of refs. \cite{adler2}, relating the trace
anomaly to the beta functions, imply that, in our case, $\Theta $ should
contain a term proportional to the Chern--Simons form, multiplied by $\beta
_{\mathrm{cs}}$. However, $\Theta $ is gauge-invariant, while the
Chern--Simons form is not. For this reason $\beta _{\mathrm{cs}}$ has to be
identically zero. This kind of argument, essentially based on the properties
of the trace anomaly, will be applied several times in this paper.

It was shown in ref. \cite{avdeev} that the Chern--Simons coupling can be
used to split the zeros of the beta function and generate a variety of
non-trivial conformal windows. For example, the beta function of a $\bar{%
\varphi}\varphi\bar{\psi}\psi$-coupling with constant $\eta $ typically
reads 
\[
\beta _{\eta }=a(\eta +bg_{\mathrm{cs}}^{2})(\eta ^{2}-g_{\mathrm{cs}}^{4}) 
\]
to the lowest order, $a$ and $b$ being some factors, possibily depending on
the gauge group and the representation. The coupling $g_{\mathrm{cs}}$ is
quantized in the non-Abelian case, $g_{\mathrm{cs}}^{2}=1/N$, and is
arbitrarily small in the large-$N$ limit. Therefore, the existence of
interacting fixed points at $\eta =\pm g_{\mathrm{cs}}^{2}$ and $\eta =-bg_{%
\mathrm{cs}}^{2}$ is proved, in this limit, directly from perturbation
theory. This construction is a three-dimensional analogue of the existence
proof of a conformal window in QCD. There we have, to two loops, 
\[
\beta _{\mathrm{QCD}}=\beta _{1}\alpha ^{2}+\beta _{2}\alpha ^{3}+\mathcal{O}%
(\alpha ^{5}),\qquad \beta _{1}=-{\frac{1}{6\pi }}(11N_{c}-2N_{f}),\qquad
\beta _{2}={\frac{25N_{c}^{2}}{(4\pi )^{2}}}, 
\]
where $\beta _{2}$ is written for $\beta _{1}\ll N_{c}$ and $N_{c}$ large.
The role of $g_{\mathrm{cs}}$ is here played by $\beta _{1}/\beta _{2}\ll
1/N_{c}$. We see that all these constructions involve a large-$N$ limit of
some sort. Our models will not be an exception in this respect.

The successful removal of divergences in quantum field theory is not
restricted to the power-counting renormalizable theories. Non-renormalizable
models in less than four dimensions were quantized long ago by Parisi, using
a large-$N$ expansion \cite{parisi}. The four-fermion model has been studied
in detail \cite{gross,rosenstein}, and the technique has been applied to
other cases, such as the $S_{N-1}$ non-linear $\sigma $-model \cite{arefeva}
and the $CP^{N-1}$ model \cite{arefeva2}. A challenging, open problem in
quantum field theory is to classify the set of power-counting
non-renormalizable theories that can be constructed in a perturbative sense,
i.e. the appropriate generalization of the power-counting criterion \cite
{me2}.

For the purposes of this paper, the Parisi large-$N$ expansion is a powerful
tool to construct non-trivial conformal field theories and conformal windows
in three dimensions. The known four-fermion models are relevant
perturbations of a certain subclass of these fixed points. Our models are
power-counting non-renormalizable, because although they do not contain
dimensionful parameters, certain bosonic fields do not have a propagator at
the classical level. Such fields are associated with composite operators and
can be scalars, but also Abelian and non-Abelian gauge vectors. The
propagators are dynamically generated by fermion loops and the large-$N$
expansion is crucial to justify the resummation of fermion bubbles before
the other diagrams, which are subleading.

To some extent, the construction presented here is a simple application of
the general theory of Parisi, however formulated in a new way, which singles
out the conformal properties and is more suitable to the research program
that we have in mind. More importantly, I generate a whole class of RG flows
(marginal deformations) interpolating between the conformal fixed points and
show that they satisfy a remarkable set of non-perturbative strong--weak
coupling dualities, also exhibited by an exact relation between the beta
function and the anomalous dimension of the composite field. The non-Abelian
gauge coupling is non-renormalized and has a discrete set of values. Observe
that our theories do not contain a Chern--Simons term. I give a general
argument proving the non-renormalization theorem, based on the trace anomaly.

I work in the Euclidean framework and use a modified
dimensional-regularization technique. The naive dimensional technique is
indeed not applicable to the theories studied here, nor to the more familiar
four-fermion models, because the dynamically generated propagator does not
regularize correctly. It is necessary to add a peculiar non-local term $%
\mathcal{L}_{\mathrm{non\,loc}}$ to the classical lagrangian. This term does
not generate new renormalization constants and is evanescent, therefore
formally absent in $D=3$.

I start from the four-fermion model, written in terms of an auxiliary field $%
\sigma$: 
\begin{equation}
\mathcal{L}_{N}=\sum_{i=1}^{N}\bar{\psi}^{i}\left( \partial \!\!\!\slash%
+\lambda \sigma \right) \psi ^{i}+\frac{1}{2}M\sigma ^{2}.  \label{poi}
\end{equation}
This theory was constructed rigorously in \cite{decalan}, where the
existence of an interacting UV\ fixed point was established. A detailed
study can be found in \cite{rosenstein}. There are two phases, and the
chiral symmetry can be dynamically broken. The $\sigma$-field equation gives 
$\sigma=-\lambda\bar{\psi}\psi/M$, whence the name ``composite boson'' for $%
\sigma$.

The theory is well-defined also if we set $M=0$. The model 
\begin{equation}
\mathcal{L}=\sum_{i=1}^{N}\bar{\psi}^{i}\left( \partial \!\!\!\slash+\lambda
\sigma \right) \psi ^{i}  \label{prima}
\end{equation}
is conformal both at the classical and quantum levels, as we now prove. We
call it the $\sigma _{N}$ conformal field theory. At the classical level no
scale is present. The renormalized lagrangian has the form 
\[
\mathcal{L}=Z_{\psi }\bar{\psi}\partial \!\!\!\slash\psi +\lambda _{\mathrm{B%
}}Z_{\sigma }^{1/2}Z_{\psi }\bar{\psi}\sigma \psi +\mathcal{L}_{\mathrm{%
non\,loc}}. 
\]
$\mathcal{L}_{\mathrm{non\,loc}}$ denotes the evanescent term to be
discussed below. No $\sigma ^{3}$-term is generated by renormalization,
because of the symmetry $x_{1}\rightarrow -x_{1}$, $\psi \rightarrow \gamma
_{1}\psi $, $\sigma \rightarrow -\sigma $. The quadratic terms in $\sigma $
are also absent: \textit{i)} the mass term $M\sigma ^{2}$ is not generated,
because it is absent in the classical lagrangian and we can choose a
subtraction scheme such that the cut-off appears only logarithmically in the
quantum action; \textit{ii)} no local kinetic term for $\sigma $ can be
generated, since the field $\sigma $ has dimension 1 in $D=3$.

In general, the bare coupling can be written as $\lambda _{\mathrm{B}%
}=\lambda Z_{\lambda }\mu ^{\varepsilon /2}$. However, the number of
independent renormalization constants is equal to the number of independent
fields and therefore we can interpret two $Z$'s as the wave-function
renormalization constants of $\psi $ and $\sigma $, and set $Z_{\lambda
}\equiv 1$. This ensures that $\beta _{\lambda }\equiv 0$ in $D=3$ and
proves that the theory is conformal also at the quantum level. At the level
of the trace anomaly, conformality (i.e. $\Theta\equiv 0$) follows from the
fact that all local dimension-3 operators are proportional to the field
equations.

The dynamical $\sigma $ kinetic term is generated by diagram (a), which,
expanded around three dimensions, gives 
\[
\mathrm{(a)}=-\frac{N\lambda _{\mathrm{B}}^{2}}{(4\pi )^{D/2}}\frac{\Gamma
\left( 2-D/2\right) \Gamma ^{2}\left( D/2-1\right) }{\Gamma \left(
D-2\right) }(k^{2})^{D/2-1}=-\frac{\lambda _{\mathrm{B}}^{2}N}{8}%
(k^{2})^{(1-\varepsilon )/2}+\mathcal{O}(\varepsilon ). 
\]
We fix the normalization with 
\begin{equation}
\lambda ^{2}N=8+\mathcal{O}(1/N),  \label{nor2}
\end{equation}
in $D=3$ and find, in momentum space, 
\begin{equation}
\Gamma _{\mathrm{kin}}[\sigma ]={\frac{1}{2}}|\sigma (k)|^{2}\mu
^{\varepsilon }(k^{2})^{(1-\varepsilon )/2}+\frac{1}{2}M\sigma ^{2}.
\label{ghu}
\end{equation}
From the diagrammatic point of view, the reader might find it easier to
imagine that the mass $M$ is still non-zero, but small, and set it to zero
at the end. In particular, at $M\neq 0$ it is immediate to resum the
geometric series of the bubbles of type (a) (see Fig. 1). After inverting
the $\sigma $ kinetic term and finding the propagator 
\begin{equation}
\langle \sigma (k)~\sigma (-k)\rangle =\frac{1}{M}\sum_{L=0}^{\infty
}(-1)^{L}\frac{\mu ^{L\varepsilon }(k^{2})^{L(1-\varepsilon )/2}}{M^{L}}=%
\frac{1}{\mu ^{\varepsilon }(k^{2})^{(1-\varepsilon )/2}+M},  \label{propp}
\end{equation}
$M$ can be freely set to $0$, which we assume from now on. We see that the
propagator of the $\sigma $-field is proportional to $1/\sqrt{k^{2}}$ in $%
D=3 $. The propagator (\ref{propp}), however, does not regularize the theory
properly, because it goes to zero too slowly at high energies. This fact
becomes apparent in the calculations of the subleading corrections. Consider
the example of diagram (b), where the dashed line is meant to be the $\sigma 
$-propagator (\ref{propp}). The integral 
\[
\int {\frac{\mathrm{d}^{3-\varepsilon }p~~~(p\!\!\!\slash+k\!\!\!\slash
)}{(p+k)^{2}(p^{2})^{(1-\varepsilon )/2}}} 
\]
produces a $\Gamma (0)$. The same holds for diagram (c). This phenomenon is
very general and concerns theories of composite bosons in every dimension,
and in particular the logarithmically trivial $D=4$ four-fermion models
considered by Wilson in \cite{wilson}. We conclude that the naive
dimensional-regularization procedure fails to regularize our theories.

\begin{figure}[tbp]
\centerline{\epsfig{figure=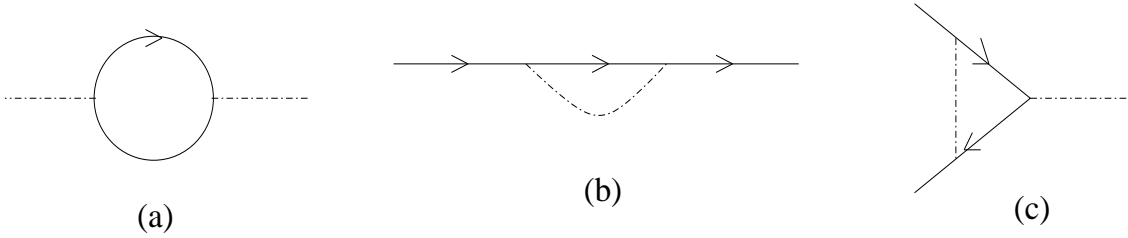,height=3cm,width=15cm}}
\caption{Leading diagram and first subleading corrections.}
\end{figure}

The problem can be cured by giving a classical, but evanescent, kinetic term
to the composite field $\sigma $, which at the leading order reads 
\begin{equation}
\mathcal{L}_{\mathrm{non\,loc}}={\frac{1}{2}}|\sigma (k)|^{2}\sqrt{k^{2}}%
\left[ 1-\frac{\lambda _{\mathrm{B}}^{2}N}{8}(k^{2})^{-\varepsilon
/2}\right] .  \label{lnonloc}
\end{equation}
$\mathcal{L}_{\mathrm{non\,loc}}$ is renormalization-group invariant. This
requirement is essential for an easier study of the theory. The new $\Gamma
_{\mathrm{kin}}$ is obtained by adding (\ref{lnonloc}) to the old one,
namely (\ref{ghu}), henceforth producing the desired high-energy behaviour: 
\begin{equation}
\Gamma _{\mathrm{kin}}^{\prime }[\varphi ]={\frac{1}{2}}|\sigma (k)|^{2}%
\sqrt{k^{2}},  \label{gaga}
\end{equation}
which regularizes the theory correctly. It is easy to go through the usual
proofs of renormalizability and locality of the counterterms with the
improved dimensional technique.

In $x$-space we find, in $D=3$, 
\begin{equation}
\langle \sigma (x)~\sigma (0)\rangle =\frac{1}{2\pi ^{2}|x|^{2}}.
\label{prop}
\end{equation}
This two-point function is intrinsically non-perturbative, since it equals
the two-point function of an elementary scalar field with anomalous
dimension $+1/2$.

The field $\psi $ has dimension $\left( D-1\right) /2$ and (\ref{lnonloc})
attributes exactly the same dimension to $\sigma $. Taking the $\mu $%
-derivative of the equation $\lambda _{\mathrm{B}}=\lambda \mu ^{\varepsilon
/2},$ we get 
\[
\beta (\lambda )=\frac{\mathrm{d}\lambda }{\mathrm{d}\ln \mu }=-\frac{%
\varepsilon }{2}\lambda . 
\]
Integrating the defining relation 
\[
\gamma _{\sigma }(\lambda )=\frac{1}{2}\frac{\mathrm{d}Z_{\sigma }(\lambda
,\varepsilon )}{\mathrm{d}\ln \mu }, 
\]
we get the $\sigma $-wave-function renormalization constant \cite
{brown,collins}: 
\[
Z_{\sigma }(\lambda ,\varepsilon )=\exp \left( -\frac{4}{\varepsilon }%
\int_{0}^{\lambda }\frac{\gamma _{\sigma }(\lambda ^{\prime })}{\lambda
^{\prime }}\mathrm{d}\lambda ^{\prime }\right) . 
\]

We assume that we work in the minimal subtraction scheme. We want to find a
closed expression for $\mathcal{L}_{\mathrm{non\,loc}}$ that properly
includes the subleading corrections. The requirements are that $\mathcal{L}_{%
\mathrm{non\,loc}}$ be renormalization-group invariant and evanescent. An
expression for $\mathcal{L}_{\mathrm{non\,loc}}$ satisfying these properties
reads, in momentum space, 
\[
\mathcal{L}_{\mathrm{non\,loc}}={\frac{1}{2}}|\sigma (k)|^{2}\sqrt{k^{2}}%
\left[ 1-\frac{\lambda _{\mathrm{B}}^{2}N}{8}(k^{2})^{-\varepsilon
/2}\right] \exp \left( \frac{4}{\varepsilon }\int_{\lambda }^{\lambda _{%
\mathrm{B}}(k^{2})^{-\varepsilon /4}}\frac{\gamma _{\sigma }(\lambda
^{\prime })}{\lambda ^{\prime }}\mathrm{d}\lambda ^{\prime }\right) . 
\]
This formula is essentially unique, the alternatives differring by scheme
redefinitions. Renormalization-group invariance is exhibited by rewriting $%
\mathcal{L}_{\mathrm{non\,loc}}$ as 
\[
\mathcal{L}_{\mathrm{non\,loc}}={\frac{1}{2}}|\sigma _{\mathrm{B}}(k)|^{2}%
\sqrt{k^{2}}\left[ 1-\frac{\lambda _{\mathrm{B}}^{2}N}{8}(k^{2})^{-%
\varepsilon /2}\right] \exp \left( \frac{4}{\varepsilon }\int_{0}^{\lambda _{%
\mathrm{B}}(k^{2})^{-\varepsilon /4}}\frac{\gamma _{\sigma }(\lambda
^{\prime })}{\lambda ^{\prime }}\mathrm{d}\lambda ^{\prime }\right) , 
\]
where $\sigma _{\mathrm{B}}=\sigma Z_{\sigma }^{1/2}$. It is easy to prove
that $\mathcal{L}_{\mathrm{non\,loc}}$ is zero in $D=3$. Indeed, we have in
the $\varepsilon \rightarrow 0$ limit:

\[
\mathcal{L}_{\mathrm{non\,loc}}\sim {\frac{1}{2}}|\sigma (k)|^{2}\sqrt{k^{2}}%
\left[ 1-\frac{\lambda _{\mathrm{B}}^{2}N}{8}(k^{2})^{-\varepsilon
/2}\right] \left( \frac{\mu ^{2}}{k^{2}}\right) ^{\gamma _{\sigma }(\lambda
)}\rightarrow 0. 
\]

A straightforward application of the Callan--Symanzik equations shows that
the $\sigma $-two-point function has the form 
\begin{equation}
\Gamma _{\sigma \sigma }=A(\lambda )\sqrt{k^{2}}\left( \frac{\mu ^{2}}{k^{2}}%
\right) ^{\gamma _{\sigma }(\lambda )},  \label{conf1}
\end{equation}
or, in $x$-space, 
\begin{equation}
\langle \sigma (x)~\sigma (0)\rangle =\frac{A^{\prime }(\lambda )}{%
|x|^{2+2\gamma _{\sigma }(\lambda )}\mu ^{2\gamma _{\sigma }(\lambda )}}.
\label{conf2}
\end{equation}
The numerical coefficients $A(\lambda )$ and $A^{\prime }(\lambda )$ do not
have here a direct physical meaning, because they are scheme-dependent and
can be changed by redefining $\mu $.

Formulas (\ref{conf1}) and (\ref{conf2}) have the expected form for a
conformal field theory. A non-vanishing anomalous dimension $\gamma _{\sigma
}(\lambda )$ proves that the theory is interacting. We now calculate $\gamma
_{\sigma }(\lambda )$ to the lowest order. We find, from diagrams (b) and
(c), 
\begin{equation}
Z_{\psi }=1-\frac{\lambda ^{2}}{6\pi ^{2}\varepsilon },\qquad \qquad
Z_{\sigma }=1+\frac{4\lambda ^{2}}{3\pi ^{2}\varepsilon }  \label{zeta}
\end{equation}
and the anomalous dimensions are 
\[
\gamma _{\psi }=\frac{2}{3N\pi ^{2}},\qquad \qquad \gamma _{\sigma }=-\frac{%
16}{3N\pi ^{2}}. 
\]
These values are in agreement with the calculations of \cite{rosenstein}
(they can be checked using the formulas (2.35a-b) of that paper, after
replacing $N$ with $N/2,$ since the authors of \cite{rosenstein} use
doublets of complex spinors). Higher-order corrections have
been studied by Gracey in refs. \cite{gracey}.
It is important to remark that $\gamma
_{\sigma }$ is negative. A negative anomalous dimension for the composite
boson is not in contradiction with unitarity. We have already observed that
the uncorrected $\sigma $-dimension is 1/2-larger than the minimum. The
unitarity bound is therefore $d_{\sigma }=1+\gamma _{\sigma }>1/2$ or $%
\gamma _{\sigma }>-1/2$, so that $\gamma _{\sigma }$ is allowed to have
negative values in three dimensions. Observe that $\gamma _{\psi }$ is
instead positive and could not be otherwise for a similar reason. To the
first subleading order we have therefore the $x$-space correlator 
\[
\langle \sigma (x)~\sigma (0)\rangle =\frac{1}{2\pi ^{2}|x|^{2-32/(3N\pi
^{2})}}. 
\]

Summarizing, we have formulated, via a large-$N$ expansion and an improved
dimensional-regularization technique, a class of interacting conformal field
theories in three dimensions. These theories are in general strongly
coupled. They become weakly coupled for $N$ large, and free for $N=\infty $.

Now, we want to define renormalization-group flows interpolating between the 
$\sigma _{N+M}$ and the $\sigma _{N}$ conformal field theories. Let us
consider the lagrangian 
\begin{equation}
\mathcal{L}_{NM}=\sum_{i=1}^{M}\bar{\chi}^{i}\left( \partial \!\!\!\slash%
+g\sigma \right) \chi ^{i}+\sum_{i=1}^{N}\bar{\psi}^{i}\left( \partial \!\!\!%
\slash+\lambda \sigma \right) \psi ^{i}.  \label{NM}
\end{equation}
Here we expand perturbatively in $g$, or actually $\bar{g}=g/\lambda $. For $%
\bar{g}=0$ we have the $\sigma _{N}$ model plus $M$ free fermions. For $\bar{%
g}=1$ we have the $\sigma _{N+M}$ model. It is therefore natural to expect
that the coupling $\bar{g}$ interpolates between the two fixed points. We
can show that there is a non-trivial beta function by studying the first
perturbative corrections. We combine the small-$\bar{g}$ expansion with the
large-$N$ expansion. We also assume that $\bar{g}^{2}M/N\ll 1$. Since $\bar{g%
}$ varies from 0 to 1, this means that $M$ is much smaller than $N$. The
renormalized lagrangian reads 
\[
\mathcal{L}_{R}=\sum_{i=1}^{M}Z_{\chi }\bar{\chi}^{i}\left( \partial \!\!\!%
\slash+gZ_{\bar{g}}Z_{\sigma }^{1/2}\sigma \right) \chi
^{i}+\sum_{i=1}^{N}Z_{\psi }\bar{\psi}^{i}\left( \partial \!\!\!\slash%
+\lambda Z_{\sigma }^{1/2}\sigma \right) \psi ^{i}+\mathcal{L}_{\mathrm{%
non\,loc}}
\]
and the evanescent, renormalization-group invariant, non-local kinetic term
reads, in the general case: 
\[
\mathcal{L}_{\mathrm{non\,loc}}={\frac{1}{2}}|\sigma (k)|^{2}\sqrt{k^{2}}%
\left[ 1-\frac{\lambda _{\mathrm{B}}^{2}N}{8}(k^{2})^{-\varepsilon
/2}\right] \exp \left( -2\int_{\ln \mu }^{\ln \sqrt{k^{2}}}\gamma _{\sigma
}(\ln \mu ^{\prime })\,\mathrm{d}\ln \mu ^{\prime }\right) .
\]
From the results (\ref{zeta}) we easily get, to the lowest order, 
\[
Z_{\psi }=1-\frac{4\mu ^{-\varepsilon }}{3N\pi ^{2}\varepsilon },\qquad
Z_{\chi }=1-\frac{4\bar{g}^{2}\mu ^{-\varepsilon }}{3N\pi ^{2}\varepsilon }%
,\qquad Z_{\sigma }=1+\frac{32\mu ^{-\varepsilon }}{3N\pi ^{2}\varepsilon }%
,\qquad Z_{\bar{g}}=1+\frac{16\left( \bar{g}^{2}-1\right) \mu ^{-\varepsilon
}}{3N\pi ^{2}\varepsilon }
\]
We therefore obtain 
\begin{equation}
\beta _{\bar{g}}=\frac{16}{3N\pi ^{2}}\bar{g}\left( \bar{g}^{2}-1\right) +%
\mathcal{O}(\bar{g}/N^{2},\bar{g}^{5}/N)  \label{lobeta}
\end{equation}
and conclude that the $\sigma _{N}$ model plus $M$ decoupled fermions is the
UV limit of the flow and the $\sigma _{N+M}$ point is the IR limit.
Remarkably, the first orders in $\bar{g}$ single out correctly both fixed
points. This means that, presumably, every truncation of the perturbative
expansion of $\beta _{\bar{g}}$\ factorizes the expected $\bar{g}\left( \bar{%
g}^{2}-1\right) $. We show below that this is indeed the case. The theories
with couplings $\bar{g}$ and $-\bar{g}$ are clearly equivalent.

The flows (\ref{NM}) satisfy a natural strong--weak coupling duality,
associated with the replacement $\bar{g}\leftrightarrow 1/\bar{g},$ $%
N\leftrightarrow M$. The dual flow interpolates from the UV  $\sigma _{M}$
model with $N$ free fermions to the IR $\sigma ^{N+M}$ model. Pairs of dual
flows have the IR limits in common. Finally, the self-dual flow has $N=M$.
We immediately realize that the $\sigma _{M}$ model plus $N$ free fermions
is the fixed point at $\bar{g}=\infty $. It is natural to conjecture that
the points $\bar{g}=0,1,\infty $ are all the fixed points of the exact beta
function. The dual flows are plotted in Fig. 2.

\begin{figure}[tbp]
\centerline{\epsfig{figure=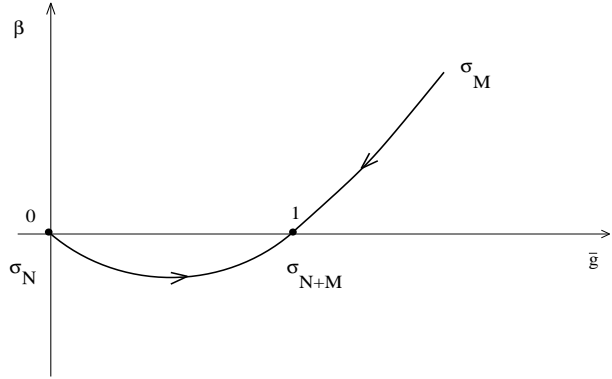,height=5cm,width=8cm}}
\caption{Beta function of dual RG flows.}
\end{figure}

The mentioned duality and fixed points are non-perturbative properties of
the flows and are self-evident from the construction. We have already seen
that, unexpectedly, the lowest order beta function (\ref{lobeta}),
calculated for $\bar{g}\ll 1$, vanishes at $\bar{g}=1$. What is even more
astonishing is that, with a little improvement, the beta function vanishes
also at $\bar{g}=\infty $ and satisfies the mentioned duality exactly. To
see this, let us relax the assumption $\bar{g}^{2}M/N\ll 1$, so that $M$ and 
$N$ can be of the same order. Diagram (a) is proportional to $N+M\bar{g}%
^{2}. $ The above formulas can be corrected replacing $N$ by $N+M\bar{g}^{2}$%
. In particular, the lowest-order beta function (\ref{lobeta}) becomes 
\[
\beta _{\bar{g}}=\frac{16}{3\pi ^{2}}{\frac{\bar{g}\left( \bar{g}%
^{2}-1\right) }{\left( N+M\bar{g}^{2}\right) }}, 
\]
and does satisfy the $\bar{g}\leftrightarrow 1/\bar{g},$ $N\leftrightarrow M$
duality, because 
\[
\beta _{1/\bar{g}}=\frac{16}{3\pi ^{2}}{\frac{1/\bar{g}\left( 1/\bar{g}%
^{2}-1\right) }{\left( M+N/\bar{g}^{2}\right) }}. 
\]
The remarkable perturbative features that we have outlined are explained by
an exact relation between the beta function and the anomalous dimension of $%
\sigma $, that we now derive. This formula is a sort of three-dimensional
analogue of certain common formulas in four-dimensional supersymmetric
theories, such as the NSVZ beta function \cite{nsvz}, or the beta function
of the superpotential coupling. We stress that in three dimensions we do not
need supersymmetry for this.

We write the renormalized lagrangian in a manifestly dual form: 
\begin{eqnarray*}
\mathcal{L}_{R} &=&\sum_{i=1}^{M}V(g,M;\lambda ,N;\varepsilon )\bar{\chi}%
^{i}\left( \partial \!\!\!\slash+gU(g,M;\lambda ,N;\varepsilon )\sigma
\right) \chi ^{i} \\
&&+\sum_{i=1}^{N}V(\lambda ,N;g,M;\varepsilon )\bar{\psi}^{i}\left( \partial
\!\!\!\slash+\lambda U(\lambda ,N;g,M;\varepsilon )\sigma \right) \psi ^{i}.
\end{eqnarray*}
We have 
\begin{eqnarray*}
Z_{\chi } &=&V(g,M;\lambda ,N;\varepsilon ),\qquad Z_{\psi }=V(\lambda
,N;g,M;\varepsilon ), \\
Z_{\sigma } &=&U^{2}(\lambda ,N;g,M;\varepsilon ),\qquad Z_{g}=\frac{%
U(g,M;\lambda ,N;\varepsilon )}{U(\lambda ,N;g,M;\varepsilon )},
\end{eqnarray*}
and find 
\begin{equation}
\beta _{\bar{g}}=\bar{g}\left( \gamma _{\sigma }-\tilde{\gamma}_{\sigma
}\right) \equiv \bar{g}\left[ \gamma _{\sigma }(\bar{g},M;N)-\gamma _{\sigma
}(1/\bar{g},N;M)\right] .  \label{beta1}
\end{equation}
Observe that $\gamma _{\sigma }(1,M;N)=\gamma _{\sigma }(M+N).$ We can
immediately check the duality of the exact beta function: 
\begin{equation}
\beta _{1/\bar{g}}=\frac{1}{\bar{g}}\left[ \gamma _{\sigma }(1/\bar{g}%
,N;M)-\gamma _{\sigma }(\bar{g},M;N)\right] .  \label{beta2}
\end{equation}
The beta function vanishes at the fixed points $\bar{g}=0,\infty $ and the
solutions of 
\begin{equation}
\gamma _{\sigma }(\bar{g},M;N)=\gamma _{\sigma }(1/\bar{g},N;M).
\label{gammagamma}
\end{equation}
Using the fact that $\gamma _{\sigma }(1,M;N)=\gamma _{\sigma }(M+N)$ we
know that $\bar{g}=1$ is a solution. We expect that this is the unique
solution of the condition (\ref{gammagamma}).

The trace anomaly reads 
\[
\Theta =\beta _{\bar{g}}\sigma \sum_{i=1}^{M}\bar{\chi}^{i}\chi ^{i}\equiv
\beta _{\bar{g}}\mathcal{O} 
\]
and, correctly, does not vanish using the field equations.

More generally, we can consider the model 
\[
\mathcal{L}=\sum_{i=1}^{k}\sum_{j=1}^{N_{i}}V(\lambda _{i},N_{i};\lambda
,N;\varepsilon )\bar{\psi}_{(i)}^{j}\left( \partial \!\!\!\slash+\lambda
_{i}U(\lambda _{i},N_{i};\lambda ,N;\varepsilon )\sigma \right) \psi
_{(i)}^{j}, 
\]
where the argument $\lambda ,N$ in $(\lambda _{i},N_{i};\lambda
,N;\varepsilon )$ refers to the set of couples $\lambda _{j},N_{j}$ with $%
j\neq i$. Clearly, $V(\lambda _{i},N_{i};\lambda ,N;\varepsilon )$ and $%
U(\lambda _{i},N_{i};\lambda ,N;\varepsilon )$ are symmetric with respect to
the exchanges $\lambda _{j},N_{j}\leftrightarrow \lambda _{l},N_{l}$ with $%
j,l\neq i$. Choosing $\lambda _{k}$ to be of order unity and all the other $%
\lambda $'s small, we have 
\begin{eqnarray*}
\beta _{i} &=&\bar{\lambda}_{i}\left( \gamma _{\sigma }-\tilde{\gamma}%
_{\sigma }^{(i)}\right) \\
&\equiv &\bar{\lambda}_{i}\left[ \gamma _{\sigma }(\bar{\lambda}%
_{1},N_{1};\cdots \lambda _{i},N_{i}\cdots ;\bar{\lambda}%
_{k-1},N_{k-1};N_{k})\right. \\
&&\left. -\gamma _{\sigma }(\bar{\lambda}_{1}/\bar{\lambda}_{i},N_{1};\cdots
1/\bar{\lambda}_{i},N_{k}\cdots ;\bar{\lambda}_{k-1}/\bar{\lambda}%
_{i},N_{k-1};N_{i})\right] ,\qquad \\
\gamma _{\sigma } &=&\gamma _{\sigma }(\bar{\lambda}_{1},N_{1};\cdots ;\bar{%
\lambda}_{k-1},N_{k-1};N_{k})=\frac{\mathrm{d}U(\lambda _{k},N_{k};\lambda
,N;\varepsilon )}{\mathrm{d}\ln \mu },
\end{eqnarray*}
where $\bar{\lambda}_{i}=\lambda _{i}/\lambda _{k}$ and $i=1,\cdots ,k-1$.
The list of fixed points is obtained by assigning the values $0,1,\infty $
to $\bar{\lambda}_{1},\ldots ,\bar{\lambda}_{k-1}$ in all possible ways,
keeping in mind that when some $\bar{\lambda}$'s are infinite, it is
immaterial whether the finite $\bar{\lambda}$'s are 0 or 1. In total, we
have $2^{k}-1$ fixed points, corresponding to the models $\sigma
_{\sum_{s}N} $ for all possible subsets $s$ of $(N_{1},\ldots ,N_{k})$. The
flows are naturally associated with the regular polyhedron having $k$ faces
in $k-1$ dimensions (triangle for $k=3$, tetrahedron for $k=4$, etc.) and
the dualities are symmetries of this polyhedron. The RG patterns for $k=2,3,4
$ are illustrated in Fig. 3.

\begin{figure}[tbp]
\centerline{\epsfig{figure=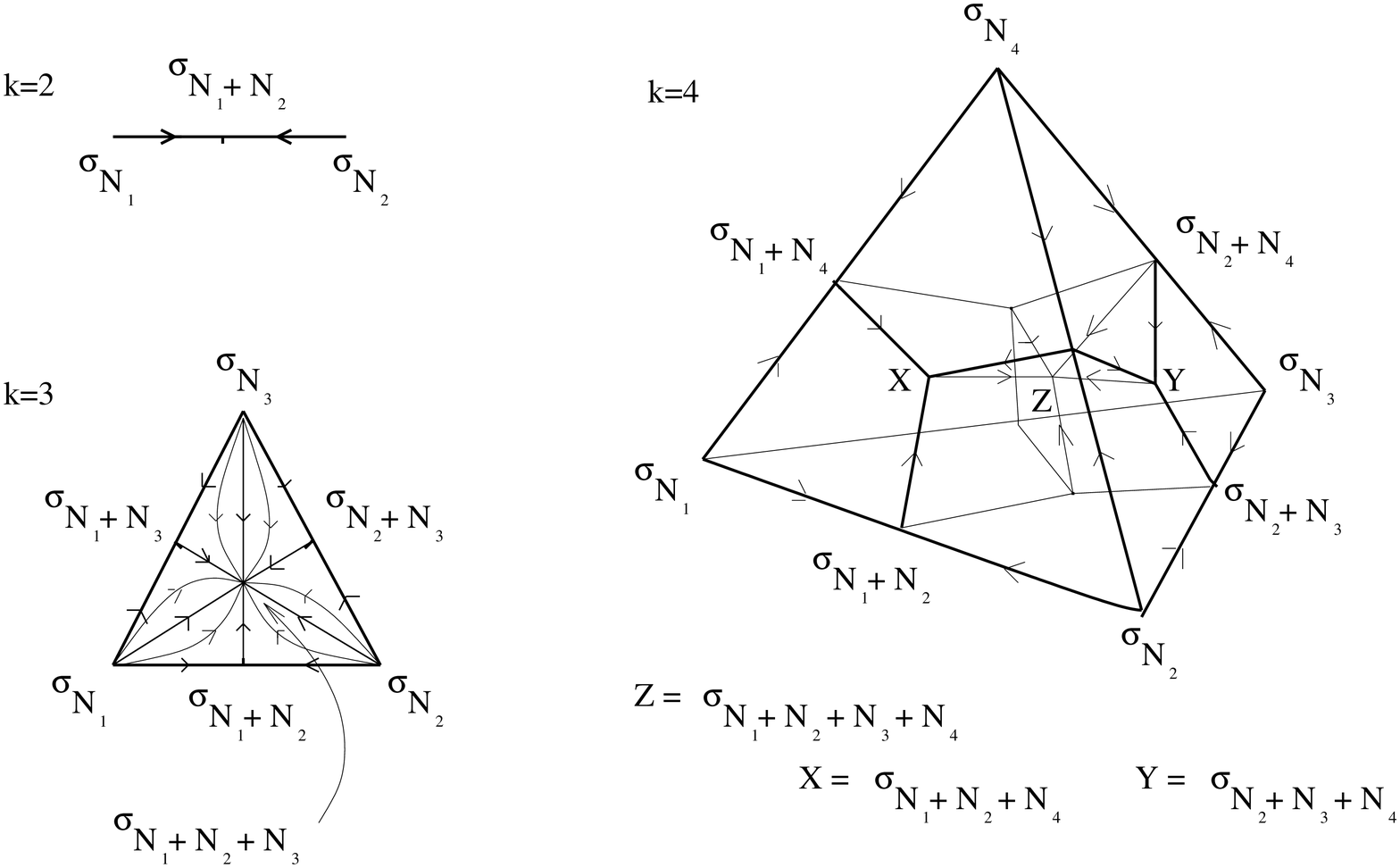,height=7.5cm,width=12cm}}
\caption{Polyhedrical RG patterns interpolating between pairs of $\sigma_N$
models.}
\end{figure}

The trace anomaly reads 
\[
\Theta=\sigma\sum_{i=1}^{k-1}\beta_i \sum_{j=1}^{N_i}\bar{\psi}%
^j_{(i)}\psi^j_{(i)}. 
\]

Flows interpolating between the UV $\sigma _{N+M}$ and IR $\sigma _{N}$
fixed points can be obtained by giving mass to $M$ fermions. For the general
purposes mentioned in the introduction, these flows are less interesting
than the pure\ RG flows, which preserve conformality at the classical level
and run only due to the dynamical scale $\mu $ \cite{cea}. In some cases,
nevertheless, such as the vector models constructed below, giving masses to
the fermions seems the only simple way to interpolate between pairs of fixed
points, because a non-renormalization theorem forbids the running of the
gauge coupling constant.

Vector four-fermion models were also considered in \cite{parisi}. Here I
make a set of observations on the non-Abelian composite gauge bosons, and
prove that their coupling constant is quantized and non-renormalized. The
Abelian coupling, instead, is non-renormalized, but can take arbitrary
values.

We start from the four-fermion model defined by the lagrangian 
\begin{equation}
\mathcal{L}=\overline{\psi }^{i}\partial \!\!\!\slash\psi ^{i}+\frac{\lambda
^{2}}{2M}\left[ (\overline{\psi }^{i}\gamma _{\mu }\psi ^{i})^{2}\right] ,
\label{serf}
\end{equation}
to which we associate the conformal field theory 
\[
\mathcal{L}=\bar{\psi}^{i}\left( \partial \!\!\!\slash+i\lambda A\!\!\!\slash%
\right) \psi ^{i}. 
\]
The vector $A_{\mu }$ becomes dynamical at the quantum level and the
resulting conformal theory is interacting. The diagram (a)\ generates the $%
A_{\mu }$-propagator at the leading order in the large-$N$ expansion and its
kinetic term in the quantum action reads, in momentum space and coordinate
space, respectively: 
\begin{equation}
\Gamma _{\mathrm{kin}}[A]=\frac{1}{2}\frac{\lambda ^{2}N}{16}A_{\mu
}(k)A_{\nu }(-k)\frac{k^{2}\delta _{\mu \nu }-k_{\mu }k_{\nu }}{\sqrt{k^{2}}}%
=\frac{1}{4}\frac{\lambda ^{2}N}{16}F_{\mu \nu }\frac{1}{\sqrt{-\Box }}%
F_{\mu \nu },  \label{quadra}
\end{equation}
$F_{\mu \nu }$ denoting the field strength. At the leading order we set
again 
\[
\frac{\lambda ^{2}N}{16}=1,\qquad \qquad \lambda _{\mathrm{B}}=\lambda \mu
^{\varepsilon /2}. 
\]
Since the $U(1)$ currents are conserved, the subleading corrections can only
change the coefficient of the quadratic term in (\ref{quadra}), but cannot
change the dimension of the vector. There is, nevertheless, a non-vanishing
anomalous dimension for the fermion fields, calculable from diagram (b) or,
alternatively, from (c). We find, using an analogue of the Feynman gauge, 
\[
\gamma _{\psi }=\frac{2}{3N\pi ^{2}}+\mathcal{O}\left( \frac{1}{N^{2}}%
\right) ,\qquad \qquad \gamma _{A}\equiv 0. 
\]
Finally, the non-local lagrangian kinetic term of the vector reads 
\[
\mathcal{L}_{\mathrm{non\,loc}}={\frac{1}{2}}A_{\mu }(k)A_{\nu }(-k)\frac{%
k^{2}\delta _{\mu \nu }-k_{\mu }k_{\nu }}{\sqrt{k^{2}}}\left[ 1-\frac{%
\lambda _{\mathrm{B}}^{2}N}{16}(k^{2})^{-\varepsilon /2}\right] 
\]
and does not need subleading corrections, since $\gamma _{A}=0$.

It is straightforward to construct conformal field theories with non-Abelian
gauge fields, using the same method. The lagrangian 
\[
\mathcal{L}=\bar{\psi}^{i}\left[ \delta _{ij}\partial \!\!\!\slash+i\lambda
A\!\!\!\slash^{a}T_{ij}^{a}\right] \psi ^{j} 
\]
generates a gauge-field quantum action 
\[
\Gamma _{\mathrm{kin}}[A]=\frac{1}{4}\frac{\lambda ^{2}C(T)N_{f}}{16}F_{\mu
\nu }^{a}\frac{1}{\sqrt{-\Box }}F_{\mu \nu }^{a}, 
\]
for the field strength 
\[
F_{\mu \nu }^{a}=\partial _{\mu }A_{\nu }^{a}-\partial _{\nu }A_{\mu
}^{a}-\lambda f^{abc}A_{\mu }^{b}A_{\nu }^{c}. 
\]
With a natural normalization convention for the action we see that the gauge
coupling is discretized and equals 
\[
g^{2}={\frac{16}{C(T)N_{f}}+}\mathcal{O}\left( \frac{1}{N_{f}^{2}}\right) . 
\]
We have the freedom to change the representation $T$ and $N_{f}$, but since
the set of unitary representations is denumerable, the non-Abelian gauge
coupling can take only discrete values. It remains to be seen whether we can
interpolate between two models with different values of the gauge coupling
by means of an RG flow. However, a non-renormalization theorem forbids this.
Consider the trace anomaly $\Theta $. The term responsible for the running
of the gauge coupling should be proportional to the gauge beta function
multiplied by a non-trivial dimension-3, local, gauge-invariant operator.
However, there exists no such operator in the gauge sector, all candidates
being proportional to the field equations. The usual term $F_{\mu \nu }^{2}$
has dimension 4, while terms proportional to $D\!\!\!\!\slash\psi $ are
trivial.

Other interesting conformal field theories are given by the gauged $\sigma
_{N}$ models 
\[
\mathcal{L}=\bar{\psi}^{i}\left[ \delta _{ij}\partial \!\!\!\slash+i\lambda
A\!\!\!\slash^{a}T_{ij}^{a}+i\lambda ^{\prime }\sigma \delta _{ij}\right]
\psi ^{j}. 
\]
RG flows such as 
\[
\mathcal{L}=\sum_{i=1}^{N}\bar{\psi}^{i}\left[ \delta _{ij}\partial \!\!\!%
\slash+i\lambda A\!\!\!\slash^{a}T_{ij}^{a}+i\lambda ^{\prime }\sigma \delta
_{ij}\right] \psi ^{j}+\sum_{i=1}^{M}\bar{\chi}^{i}\left[ \delta
_{ij}\partial \!\!\!\slash+i\lambda A\!\!\!\slash^{a}R_{ij}^{a}+ig\sigma
\delta _{ij}\right] \chi ^{j} 
\]
do not change the gauge-coupling, by the non-renormalization theorem proved
above, but only the $\sigma $ coupling. The patterns of their RG flows are
similar to the RG\ patterns of the non-gauged $\sigma _{N}$ models, with the
only difference that the duality symmetries involve also exchanges of the
representations, such as $R\leftrightarrow T$, etc.

A non-Abelian coupling can take arbitrary values, but the
non-renormalization theorem applies. Consider for example 
\[
\mathcal{L}=\sum_{i=1}^{N}\bar{\psi}^{i}\left[ \partial \!\!\!\slash%
+i\lambda A\!\!\!\slash
\right] \psi ^{i}+\sum_{i=1}^{M}\bar{\chi}^{i}\left[ \partial \!\!\!\slash%
+i\lambda ^{\prime }A\!\!\!\slash
\right] \chi ^{i}. 
\]
Here the trace anomaly is still identically zero and the theory is conformal
for arbitrary $\lambda $ and $\lambda ^{\prime }$. We have 
\[
\Gamma _{\mathrm{kin}}[A]=\frac{1}{4}\frac{\lambda ^{2}N+\lambda ^{\prime
}{}^{2}M}{16}F_{\mu \nu }\frac{1}{\sqrt{-\Box }}F_{\mu \nu }. 
\]

We can reduce to the original vector four-fermion model (\ref{serf}) by
means of a relevant deformation. A mass perturbation, such as $MA_{\mu
}^{2}/2$, produces the propagator 
\[
\langle A_{\mu }^{a}(k)~A_{\nu }^{b}(-k)\rangle =\frac{\delta ^{ab}\sqrt{%
k^{2}}}{k^{2}+M\sqrt{k^{2}}}\left( \delta _{\mu \nu }+\frac{k_{\mu }k_{\nu }%
}{M\sqrt{k^{2}}}\right) . 
\]
The behaviour $k_{\mu }k_{\nu }/(Mk^{2})$ at large momenta is not dangerous
if the current is conserved \cite{parisi}, which happens for Abelian fields.
The situation is similar to quantum electrodynamics in four dimensions,
where the photon can be given a mass without spoiling the renormalizability.
With non-Abelian gauge fields we have to advocate a symmetry-breaking
mechanism. We consider 
\[
\mathcal{L}=\bar{\psi}^{i}\left( \delta _{ij}\partial \!\!\!\slash 
+i\lambda A\!\!\!\slash^{a}T_{ij}^{a}\right) \psi ^{j}+\left| D_{\mu
}\varphi \right| ^{2}+V(|\varphi |)+\Lambda \bar{\psi}\psi \,\bar{\varphi}%
\varphi +\Lambda ^{\prime }\bar{\psi}T^{a}\psi \,\bar{\varphi}R^{a}\varphi
\cdots 
\]
and assume that the potential $V(|\varphi |)$ is such that the scalar field
has an expectation value $\langle |\varphi |\rangle =M^{1/2}$. We know that
the theory is renormalizable in the large-$N$ expansion. We can integrate
the vector field out by solving its field equation. For simplicity we write
the formulas in the Abelian case, although the mechanism is not strictly
necessary there. We have 
\[
A_{\mu }^{a}=-\frac{i}{2\lambda \left| \varphi \right| ^{2}}\left( \overline{%
\psi }\gamma _{\mu }\psi -\bar{\varphi}\partial _{\mu }\varphi +\partial
_{\mu }\varphi \bar{\varphi}\right) =-\frac{i\overline{\psi }\gamma _{\mu
}\psi }{\lambda |M^{1/2}+\eta |^{2}}-\frac{1}{\lambda }\partial _{\mu
}\theta 
\]
having set $\varphi =\mathrm{e}^{i\theta }\left( M^{1/2}+\eta \right) /\sqrt{%
2}$. The Goldstone boson $\theta $ is gauged away as usual and we remain
with 
\begin{equation}
\mathcal{L}=\bar{\psi}^{i}\partial \!\!\!\slash\psi ^{i}-\frac{\left( 
\overline{\psi }^{i}\gamma _{\mu }\psi ^{i}\right) ^{2}}{2M\left| 1+\eta /%
\sqrt{M}\right| ^{2}}+\frac{1}{2}\left( \partial _{\mu }\eta \right)
^{2}+V(\eta )+\frac{\Lambda }{2}\bar{\psi}\psi \left| 1+\eta /\sqrt{M}%
\right| ^{2}.  \label{teor}
\end{equation}
When the mass of $\eta $ is very large, (\ref{serf}) is recovered exactly.
By construction, the theory (\ref{teor}) is renormalizable, although this is
not evident in the final form. Since the limit of large $\eta $-mass can be
taken at $M$ fixed, we see that (\ref{serf}) is also renormalizable.

A more direct way to get to (\ref{serf}) is by replacing $V(|\varphi |)$
with $i\alpha (\bar{\varphi}\varphi -M)$, such as in the $S_{N-1}$
non-lineal $\sigma $-model \cite{arefeva}. The field $\alpha $ is dynamical
and acquires a propagator proportional to $\sqrt{k^{2}}$, which is however
compatible with power counting. In this case, however, we need to take a
large-$N$ limit also in the number of $\varphi $ components.

\vskip 1truecm \textbf{Acknowledgements} \vskip .2truecm

I am grateful to F. Nogueira for drawing my attention to the properties of
four-fermion theories in three dimensions and the large-$N$ expansion for
non-renormalizable theories, for discussions, and for pointing out relevant
references. I also thank C. De Calan for discussions and U. Aglietti for
continuous assistance and critical remarks. Finally, I thank J. Gracey and
A. Kapustin and for pointing out references \cite{gracey} and \cite
{nonre2,kasputin}, respectively.


\begin{thebibliography}{99}
\bibitem{adler}  S. Adler, Axial-vector vertex in spinor electrodynamics,
Phys. Rev. 177 (1969) 2426.

\bibitem{jackiw}  J.S. Bell and R. Jackiw, A PCAC puzzle $\pi
^{0}\rightarrow \gamma \gamma $ in the $\sigma $-model, Nuovo Cim. A 60
(1969) 47.

\bibitem{adlerbardeen}  S.L. Adler and W.A. Bardeen, Absence of higher-order
corrections in the anomalous axial vector divergence equation, Phys. Rev.
182 (1969) 1517.

\bibitem{thooftmatching}  G. 't Hooft, in \textit{Recent developments in
gauge theories}, eds. G. 't Hooft et al. (Plenum Press, New York, 1980).

\bibitem{adler2}  S.L. Adler, J.C. Collins and A. Duncan,
Energy--momentum-tensor trace anomaly in spin 1/2 electrodynamics, Phys.
Rev. D15 (1977) 1712.

N.K. Nielsen, The energy momentum tensor in a nonabelian quark gluon theory,
Nucl. Phys. B 120 (1977) 212.

J.C. Collins, A. Duncan and S.D. Joglekar, Trace and dilatation anomalies in
gauge theories, Phys. Rev. D 16 (1977) 438.

\bibitem{noi}  D. Anselmi, D.Z. Freedman, M.T. Grisaru and A.A. Johansen,
Nonperturbative formulas for central functions in supersymmetric theories,
Nucl. Phys. B 526 (1998) 543 and hep-th/9708042.

\bibitem{noi2}  D. Anselmi, J. Erlich, D.Z. Freedman and A.A. Johansen,
Positivity constraints on anomalies in supersymmetric gauge theories, Phys.
Rev. D57 (1998) 7570 and hep-th/9711035.

\bibitem{nsvz}  V. Novikov, M.A. Shifman, A.I. Vainshtein and V. Zakharov,
Exact Gell-Mann--Low Function of Supersymmetric Yang--Mills Theories from
Instanton Calculus, Nucl. Phys. B229 (1983) 381.

\bibitem{athm}  D. Anselmi, Anomalies, unitarity and quantum
irreversibility, Ann. Phys. (NY) 276 (1999) 361 and hep-th/9903059.

\bibitem{cea}  D. Anselmi, Towards the classification of conformal field
theories in arbitrary even dimension, Phys. Lett. B 476 (2000) 182 and
hep-th/9908014.

\bibitem{at6d}  D. Anselmi, Quantum irreversibility in arbitrary dimension,
Nucl. Phys. B 567 (2000) 331 and hep-th/9905005.

\bibitem{avdeev}  L.V. Avdeev, G.V. Grigorev and D.I. Kazakov,
Renormalization in Abelian Chern--Simons field theories with matter, Nucl.
Phys. B 382 (1992) 561;

L.V. Avdeev, D.I. Kazakov and I.N. Kondrashuk, Renormalizations in
supersymmetric and nonsupersymmetric non-Abelian Chern--Simons field
theories with matter, Nucl. Phys. B 391 (1993) 333.

\bibitem{townsend} P.K. Townsend, Consistency of the $1/N$
expansion for the three-dimensional $\phi^6$ theory, Nucl. Phys. B
118 (1977) 199.

\bibitem{nonre}  A. Blasi and R. Collina, Finiteness of the Chern--Simons
model in perturbation theory, Nucl. Phys. B 345 (1990) 472;

\bibitem{nonre2}  A. Blasi, N. Maggiore and S.P. Sorella, Nonrenormalization
properties of the Chern--Simons action coupled to matter, Phys. Lett. B 285
(1992) 54 and hep-th/9204045. 

\bibitem{kasputin}  A.N. Kapustin and P.I. Pronin, Non-renormalization
theorem for the gauge coupling in 2+1 dimensions, Mod. Phys. Lett. A9 (1994)
1925 and hep-th/9401053.

\bibitem{parisi}  G. Parisi, The theory of non-renormalizable interactions.
--- I. The large-$N$ expansion, Nucl. Phys. B 100 (1975) 368.

\bibitem{gross}  D.J. Gross, \textit{Applications of the renormalization
group to high-energy physics}, in Les Houches, Session XXVIII, Methods in
Field Theory, eds. R. Balian and J. Zinn-Justin (North Holland Publishing
Company, Amsterdam, 1976).

\bibitem{rosenstein}  B. Rosenstein, B. Warr and S.H. Park, Dynamical
symmetry breaking in four-fermion interaction models, Phys. Rep. 205 (1991)
59.

\bibitem{arefeva}  I.Ya. Aref'eva, Phase transition in the three dimensional
chiral field, Ann. Phys. 117 (1979) 393.

\bibitem{arefeva2}  I.Ya. Aref'eva and S.I. Azakov, Renormalization and
phase transition in the quantum $CP^{N-1}$ model ($D=2,3$), Nucl. Phys. B
162 (1980) 298.

\bibitem{me2}  D. Anselmi, More on the subtraction algorithm, Class. and
Quantum Grav. 12 (1995) 319 and hep-th/9407023.

\bibitem{decalan}  C. de Calan, P.A. Faria da Vega, J. Magnen and R. Seneor,
Constructing the three-dimensional Gross-Neveu model with a large number of
flavour components, Phys. Rev. Lett. 66 (1991) 3233.

\bibitem{brown}  L.S. Brown, \textit{Quantum field theory}, Cambridge
University Press, Cambridge, 1992.

\bibitem{collins}  J.C. Collins, \textit{Renormalization}, Cambridge
University Press, Cambridge, 1984.

\bibitem{gracey} J.A. Gracey, Computation of $\beta'(g_c)$ at
${\cal O}(1/N^2)$ in the $O(N)$ Gross Neveu model in arbitrary
dimensions, Int. J. Mod. Phys. A9 (1994) 567 and hep-th/9306106;

Computation of critical exponent $\eta$ at ${\cal O}(1/N^3)$ 
in the four fermi model in arbitrary dimensions, Int. J. Mod.
Phys. A9 (1994) 727 and hep-th/9306107.

\bibitem{wilson}  K. Wilson, Quantum field-theory models in less than 4
dimensions, Phys. Rev. D 7 (1973) 2911.
\end{thebibliography}
\end{document}